\documentclass[aps,prd,twocolumn,showkeys,amsmath,amssymb]{revtex4}
\usepackage{graphicx}
\usepackage{epstopdf}
\usepackage{multirow}
\usepackage[colorlinks,citecolor=blue,anchorcolor=red,menucolor=red,linkcolor=red,filecolor=red,runcolor=red,urlcolor=blue,frenchlinks=red]{hyperref}
\allowdisplaybreaks[3]

\begin{document}
%=====================================================================================
%=====================================================================================
\title{Toward the existence of odderon as a three-gluon bound state}
%=====================================================================================
%=====================================================================================
%

\author{Hua-Xing Chen$^1$}
\email{hxchen@seu.edu.cn}
\author{Wei Chen$^2$}
\email{chenwei29@mail.sysu.edu.cn}
\author{Shi-Lin Zhu$^{3,4,5}$}
\email{zhusl@pku.edu.cn}

\affiliation{
$^1$School of Physics, Southeast University, Nanjing 210094, China\\
$^2$School of Physics, Sun Yat-Sen University, Guangzhou 510275, China\\
$^3$School of Physics and State Key Laboratory of Nuclear Physics and Technology, Peking University, Beijing 100871, China\\
$^4$Collaborative Innovation Center of Quantum Matter, Beijing 100871, China\\
$^5$Center of High Energy Physics, Peking University, Beijing 100871, China}

\begin{abstract}
Inspired by the evidence of the odderon exchange recently observed by the D0 and TOTEM Collaborations, a QCD sum rule investigation is performed to study the odderon as a three-gluon bound state. There may exist six lowest-lying three-gluon odderons with the quantum numbers $J^{PC} = 1/2/3^{\pm-}$. We systematically construct their interpolating currents, and calculate their mass spectra. To verify their existence, we propose to search for the spin-3 odderons in their $VVV$ and $VVP$ decay channels directly at LHC, with $V$ and $P$ light vector and pseudoscalar mesons respectively.
\end{abstract}
\maketitle
\pagenumbering{arabic}

$\\$
{\it Introduction.}-----
Very recently, the D0 and TOTEM Collaborations compared their $pp$ and $p\bar p$~\cite{Abazov:2012qb} cross sections, and found they differ with a significance of $3.4\sigma$~\cite{Abazov:2020rus}. This leads to the evidence of a $t$-channel exchanged odderon~\cite{Cudell:2002xe,Martynov:2018sga}, {\it i.e.}, a colorless $C$-odd gluonic compound. They further combined their previous result of Ref.~\cite{Antchev:2017yns}, and increased the significance of this evidence to $5.2\sigma$--$5.7\sigma$. Accordingly, the D0 and TOTEM Collaborations claimed that they have accomplished the first experimental observation of the odderon!

Actually, in the previous TOTEM experiment~\cite{Antchev:2017yns} they had attempted to probe the existence of the odderon, and there had been some discussions on it~\cite{Khoze:2018bus,Csorgo:2018uyp,Goncalves:2018nsp,Xie:2019soz}. However, these discussions are mainly on the contribution of the odderon exchange in certain collisions, so it is crucial and important to directly study the odderon itself. Recall that the odderon was first introduced in 1973~\cite{Lukaszuk:1973nt}, and later reintroduced in QCD in the 1980's~\cite{Bartels:1980pe,Kwiecinski:1980wb,Donnachie:1983ff}. There had been several attempts to prove its existence, and we refer to the review~\cite{Block:2006hy} for detailed discussions.

The odderon is probably/predominantly a $C$-odd three-gluon bound state, that is one category of glueballs. Glueballs are important for the understanding of non-perturbative QCD, but there is currently no definite
experimental evidence for their existence, although there have been tremendous of discussions on them using various theoretical methods and models, such as the MIT bag model~\cite{Chodos:1974je,Jaffe:1975fd}, the flux tube model~\cite{Isgur:1984bm}, the Coulomb Gauge model~\cite{Szczepaniak:1995cw,LlanesEstrada:2005jf}, Lattice QCD~\cite{Wilson:1974sk,Chen:2005mg,Mathieu:2008me,Meyer:2004gx,Gregory:2012hu}, and QCD sum rules~\cite{Novikov:1979va,Narison:1984hu,Narison:1996fm,Bagan:1990sy}, etc. Especially, mass spectra of the spin-0 odderons were systematically studies in Refs.~\cite{Latorre:1987wt,Hao:2005hu,Qiao:2014vva,Qiao:2015iea,Pimikov:2016pag} by using QCD sum rules. However, in Ref.~\cite{Szanyi:2019kkn} the authors studied glueball trajectories, and their results suggest that the odderon is essentially an object in the Regge pole model, and any derivation of its finite mass from massless gluons meets conceptual difficulties in QCD, preventing reliable calculations of its spectrum.

The lowest-lying two-, three-, and four-glueballs were systematically investigated in Ref.~\cite{Jaffe:1985qp}, where the authors classified altogether six lowest-lying odderons with the quantum numbers $J^{PC} = 1/2/3^{\pm-}$, and constructed their corresponding non-relativistic low-dimension operators. Some of these operators have been successfully used in Lattice QCD calculations. This inspires us to construct their corresponding relativistic odderon currents, which can be further applied to perform QCD sum rule analyses and calculate masses of the $J^{PC} = 1/2/3^{\pm-}$ odderons. This is just the aim of the present letter. Comparison of our QCD sum rule results with the Lattice QCD results~\cite{Chen:2005mg,Mathieu:2008me,Meyer:2004gx,Gregory:2012hu} can be found in Table~\ref{tab:comparison} at the end of this letter.

$\\$
{\it Relativistic odderon currents.}-----
As the first step, we use the gluon field strength tensor $G^a_{\mu\nu}$ to construct relativistic odderon currents. Here $a$ is a color index, and $\mu,\nu$ are Lorentz indices. Besides, we use $d^{abc}$ to denote the totally symmetric $SU_c(3)$ structure constant, and $\tilde G^a_{\mu\nu}$ to denote the dual gluon field strength tensor, defined as $\tilde G^a_{\mu\nu} = G^{a,\rho\sigma} \times \epsilon_{\mu\nu\rho\sigma}/2$.

In Ref.~\cite{Jaffe:1985qp} the chromoelectric and chromomagnetic fields ($i,j=1,2,3$),
\begin{equation}
E_i = G_{i0} ~~~ {\rm and} ~~~ B_i = -{1\over2} \epsilon_{ijk} G^{jk} \, ,
\end{equation}
were used to write down the non-relativistic low-dimension odderon operators:
\begin{eqnarray}
\nonumber && 1^{+-} ~~~~~ d^{abc} ( \vec E_a \cdot \vec E_b ) \vec B_c \, ,
\\ \nonumber && 1^{--} ~~~~~ d^{abc} ( \vec E_a \cdot \vec E_b ) \vec E_c \, ,
\\ \nonumber && 2^{+-} ~~~~~ d^{abc} \mathcal{S}[ E_a^i ( \vec E_b \times \vec B_c )^j ] \, ,
\\ \nonumber && 2^{--} ~~~~~ d^{abc} \mathcal{S}[ B_a^i ( \vec E_b \times \vec B_c )^j ] \, ,
\\ \nonumber && 3^{+-} ~~~~~ d^{abc} \mathcal{S}[ B_a^i B_b^j B_c^k] \, ,
\\ && 3^{--} ~~~~~ d^{abc} \mathcal{S}[ E_a^i E_b^j E_c^k] \, ,
\end{eqnarray}
where $\mathcal{S}$ denotes symmetrization and subtracting trace terms in the set $\{ij\}$ or $\{ijk\}$.

To perform QCD sum rule analyses, we further construct their corresponding relativistic currents:
\begin{eqnarray}
J_1^{\alpha\beta} &=& d^{abc} G_a^{\mu\nu} G_{b,\mu\nu} G_{c}^{\alpha\beta} \, ,
\\
\nonumber J_2^{\alpha_1\alpha_2,\beta_1\beta_2} &=& d^{abc} \mathcal{S}^\prime[ G_a^{\alpha_1\beta_1} \tilde G_b^{\alpha_2\mu} G_{c,\mu}^{\beta_2} - \{ \alpha_2 \leftrightarrow \beta_2 \} ] ,
\\
\\
J_3^{\alpha_1\alpha_2\alpha_3,\beta_1\beta_2\beta_3} &=& d^{abc} \mathcal{S}^\prime[ G_a^{\alpha_1\beta_1} G_{b}^{\alpha_2\beta_2} G_c^{\alpha_3\beta_3} ] \, ,
\end{eqnarray}
where $\mathcal{S}^\prime$ denotes symmetrization and subtracting trace terms in the two sets $\{\alpha_1 \cdots \alpha_J\}$ and $\{\beta_1 \cdots \beta_J\}$ simultaneously.

The first current $J_1^{\alpha\beta}$ contains both $J^{PC} = 1^{+-}$ and $1^{--}$ components, and it couples to the $1^{+-}$ and $1^{--}$ odderons simultaneously:
\begin{eqnarray}
\langle 0 | J_1^{\alpha\beta} | X_{1^{+-}} \rangle &=& i f_{1^{+-}} \epsilon^{\alpha\beta \mu \nu} \epsilon_\mu p_\nu \, ,
\label{eq:coupling2}
\\
\langle 0 | J_1^{\alpha\beta} | X_{1^{--}} \rangle &=& i f_{1^{--}} (p^\alpha \epsilon^\beta - p^\beta \epsilon^\alpha) \, ,
\end{eqnarray}
where $f_{1^{+-}}$ and $f_{1^{--}}$ are decay constants, and $\epsilon_\mu$ is the polarization vector. This current has a partner
\begin{equation}
\tilde J_1^{\alpha\beta} = d^{abc} \tilde G_a^{\mu\nu} \tilde G_{b,\mu\nu} \tilde G_{c}^{\alpha\beta} \, ,
\end{equation}
which just has the opposite couplings:
\begin{eqnarray}
\langle 0 | \tilde J_1^{\alpha\beta} | X_{1^{+-}} \rangle &=& i f_{1^{+-}} (p^\alpha \epsilon^\beta - p^\beta \epsilon^\alpha) \, ,
\\ \langle 0 | \tilde J_1^{\alpha\beta} | X_{1^{--}} \rangle &=& i f_{1^{--}} \epsilon^{\alpha\beta \mu \nu} \epsilon_\mu p_\nu \, .
\label{eq:coupling3}
\end{eqnarray}
Therefore, we can use either $J_1^{\alpha\beta}$ or $\tilde J_1^{\alpha\beta}$ to investigate the $1^{+-}$ and $1^{--}$ odderons at the same time. In the present study we use $J_1^{\alpha\beta}$ to investigate the $1^{+-}$ odderon and $\tilde J_1^{\alpha\beta}$ to investigate the $1^{--}$ one, because the couplings given in Eqs.~(\ref{eq:coupling2}) and (\ref{eq:coupling3}) can be more easily calculated.

The currents $J_2^{\alpha_1\alpha_2,\beta_1\beta_2}$ and its partner
\begin{equation}
\tilde J_2^{\alpha_1\alpha_2,\beta_1\beta_2} = d^{abc} \mathcal{S}^\prime[ \tilde G_a^{\alpha_1\beta_1} G_b^{\alpha_2\mu} \tilde G_{c,\mu}^{\beta_2} - \{ \alpha_2 \leftrightarrow \beta_2 \} ] ,
\end{equation}
couple to the $2^{+-}$ and $2^{--}$ odderons through:
\begin{eqnarray}
\nonumber \langle 0 | J_2^{\cdots} | X_{2^{+-}} \rangle &=& i f_{2^{+-}} \mathcal{S}^{\prime}[ \epsilon^{\alpha_i \beta_i \mu_i \nu_i} \epsilon_{\mu_i} p_{\nu_i} ]^2 \, ,
\\ \langle 0 | \tilde J_2^{\cdots} | X_{2^{--}} \rangle &=& i f_{2^{--}} \mathcal{S}^{\prime}[ \epsilon^{\alpha_i \beta_i \mu_i \nu_i} \epsilon_{\mu_i} p_{\nu_i} ]^2 \, ,
\end{eqnarray}
where
\begin{equation}
[ \cdots ]^J = \epsilon^{\alpha_1 \beta_1 \mu_1 \nu_1} \epsilon_{\mu_1} p_{\nu_1} \cdots \epsilon^{\alpha_J \beta_J \mu_J \nu_J} \epsilon_{\mu_J} p_{\nu_J} \, .
\end{equation}

The current $J_3^{\alpha_1\alpha_2\alpha_3,\beta_1\beta_2\beta_3}$ and its partner
\begin{equation}
\tilde J_3^{\alpha_1\alpha_2\alpha_3,\beta_1\beta_2\beta_3} = d^{abc} \mathcal{S}^{\prime}[\tilde G_a^{\alpha_1\beta_1} \tilde G_{b}^{\alpha_2\beta_2} \tilde G_c^{\alpha_3\beta_3}] \, ,
\end{equation}
couple to the $3^{+-}$ and $3^{--}$ odderons through:
\begin{eqnarray}
\nonumber \langle 0 | J_3^{\cdots} | X_{3^{+-}} \rangle &=& i f_{3^{+-}} \mathcal{S}^{\prime}[ \epsilon^{\alpha_i \beta_i \mu_i \nu_i} \epsilon_{\mu_i} p_{\nu_i} ]^3 \, ,
\\ \langle 0 | \tilde J_3^{\cdots} | X_{3^{--}} \rangle &=& i f_{3^{--}} \mathcal{S}^{\prime}[ \epsilon^{\alpha_i \beta_i \mu_i \nu_i} \epsilon_{\mu_i} p_{\nu_i} ]^3 \, .
\end{eqnarray}

It is interesting to notice that all the above currents have $N=2J$ Lorentz indices with certain symmetries, so that they can couple to the positive- and negative-parity odderons simultaneously. For example, the spin-2 current $J_2^{\alpha_1\alpha_2,\beta_1\beta_2}$ have four Lorentz indices, satisfying
\begin{equation}
J_2^{\alpha_1\alpha_2,\beta_1\beta_2} = - J_2^{\beta_1\alpha_2,\alpha_1\beta_2} = - J_2^{\alpha_1\beta_2,\beta_1\alpha_2} = J_2^{\alpha_2\alpha_1,\beta_2\beta_1} \, .
\end{equation}

$\\$
{\it QCD sum rule analyses.}-----
In this section we use the current $J_1^{\alpha\beta}$ as an example to perform QCD sum rule analyses, which method has been widely applied in the study of hadron phenomenology~\cite{Shifman:1978bx,Reinders:1984sr}. We study the two-point correlation function
%
%%%%%%%%%%%%%%%%%%%%%%%%%%%%%%%%%%%%%%%%%%%%%%%%%%%%%%%%%%%%%%%%%%%%%%%%%%%%%%
\begin{eqnarray}
\nonumber \Pi^{\alpha\beta,\alpha^\prime\beta^\prime}(q^2) &\equiv& i \int d^4x e^{iqx} \langle 0 | {\bf T}[J_1^{\alpha\beta}(x) J_1^{\alpha^\prime\beta^\prime\dagger}(0)] | 0 \rangle
\\ &=& (g^{\alpha\alpha^\prime}g^{\beta\beta^\prime} - g^{\alpha\beta^\prime}g^{\beta\alpha^\prime})~\Pi (q^2) \, ,
\label{def:pi}
\end{eqnarray}
%%%%%%%%%%%%%%%%%%%%%%%%%%%%%%%%%%%%%%%%%%%%%%%%%%%%%%%%%%%%%%%%%%%%%%%%%%%%%%
%
at both hadron and quark-gluon levels.

At the hadron level we use the dispersion relation to express Eq.~(\ref{def:pi}) as
%
%%%%%%%%%%%%%%%%%%%%%%%%%%%%%%%%%%%%%%%%%%%%%%%%%%%%%%%%%%%%%%%%%%%%%%%%%%%%%%
\begin{equation}
\Pi(q^2) = \int^\infty_{0}\frac{\rho(s)}{s-q^2-i\varepsilon}ds \, ,
\label{def:rho}
\end{equation}
%%%%%%%%%%%%%%%%%%%%%%%%%%%%%%%%%%%%%%%%%%%%%%%%%%%%%%%%%%%%%%%%%%%%%%%%%%%%%%
%
where $\rho(s) \equiv {\rm Im}\Pi(s)/\pi$ is the spectral density. We parameterize it using one
pole dominance for the ground state $X$ together with the continuum contribution,
%
%%%%%%%%%%%%%%%%%%%%%%%%%%%%%%%%%%%%%%%%%%%%%%%%%%%%%%%%%%%%%%%%%%%%%%%%%%%%%%
\begin{eqnarray}
\nonumber \rho(s) &\equiv& \sum_n\delta(s-M^2_n) \langle 0| J | n\rangle \langle n| J^{\dagger} |0 \rangle
\\ &=& f^2_X \delta(s-M^2_X) + \rm{continuum} \, .
\label{eq:rho}
\end{eqnarray}
%%%%%%%%%%%%%%%%%%%%%%%%%%%%%%%%%%%%%%%%%%%%%%%%%%%%%%%%%%%%%%%%%%%%%%%%%%%%%%
%

At the quark-gluon level we insert $J_1^{\alpha\beta}$ into Eq.~(\ref{def:pi}) and calculate it using the method of operator product expansion (OPE). Then we perform the Borel transformation to Eq.~(\ref{def:rho}) at both hadron and quark-gluon levels. After approximating the continuum using the spectral density above a threshold value $s_0$, we obtain
%
%%%%%%%%%%%%%%%%%%%%%%%%%%%%%%%%%%%%%%%%%%%%%%%%%%%%%%%%%%%%%%%%%%%%%%%%%%%%%%
\begin{equation}
\Pi(s_0, M_B^2) \equiv f^2_X e^{-M_X^2/M_B^2} = \int^{s_0}_{0} e^{-s/M_B^2}\rho(s)ds \, ,
\label{eq:fin}
\end{equation}
%%%%%%%%%%%%%%%%%%%%%%%%%%%%%%%%%%%%%%%%%%%%%%%%%%%%%%%%%%%%%%%%%%%%%%%%%%%%%%
%
which can be used to calculate the mass of $X$ through
%
%%%%%%%%%%%%%%%%%%%%%%%%%%%%%%%%%%%%%%%%%%%%%%%%%%%%%%%%%%%%%%%%%%%%%%%%%%%%%%
\begin{equation}
M^2_X(s_0, M_B) = \frac{\int^{s_0}_{0} e^{-s/M_B^2}s\rho(s)ds}{\int^{s_0}_{0} e^{-s/M_B^2}\rho(s)ds} \, .
\label{eq:LSR}
\end{equation}
%%%%%%%%%%%%%%%%%%%%%%%%%%%%%%%%%%%%%%%%%%%%%%%%%%%%%%%%%%%%%%%%%%%%%%%%%%%%%%
%

In the present study we calculate OPEs up to the dimension eight ($D=8$) condensates, including the perturbative term, the two-gluon condensate $\langle g_s^2 GG \rangle \equiv \langle g_s^2 G^a_{\mu\nu} G_a^{\mu\nu} \rangle$, the three-gluon condensate $\langle g_s^3 G^3 \rangle \equiv \langle g_s^3 f^{abc} G_a^{\mu\nu} G_{b,\nu\rho} G_{c,\mu}^{\rho} \rangle$, and their combination $\langle g_s^2 GG \rangle^2$:
%%%%%%%%%%%%%%%%%%%%%%%%%%%%%%%%%%%%%%%%%%%%%%%%%%%%%%%%%%%%%%%%%%%%%%%%%%%%%%
\begin{eqnarray}
%------------------------------\rho 1-- eta_1----------------------------------
\nonumber \rho_{1^{+-}}(s) &=& \frac{4 \alpha_s^3}{81 \pi} s^4 + \frac{10\alpha_s^2 \langle g_s^2 GG \rangle}{9} s^2 + \frac{35\alpha_s^3 \langle g_s^2 GG \rangle}{36\pi} s^2
\\ && - \frac{205\alpha_s^2 \langle g_s^3 G^3 \rangle}{54} s \, ,
\label{eq:ope1pm}
\\
\nonumber \rho_{1^{--}}(s) &=& \frac{4 \alpha_s^3}{81 \pi} s^4 - \frac{10\alpha_s^2 \langle g_s^2 GG \rangle}{9} s^2 + \frac{25\alpha_s^3 \langle g_s^2 GG \rangle}{36\pi} s^2
\\ && + \frac{5\alpha_s^2 \langle g_s^3 G^3 \rangle}{54} s \, ,
\label{eq:ope1mm}
\\
\nonumber \rho_{2^{+-}}(s) &=& \frac{\alpha_s^3}{81 \pi} s^4 + \frac{5\alpha_s^2 \langle g_s^2 GG \rangle}{27} s^2 + \frac{15\alpha_s^3 \langle g_s^2 GG \rangle}{32\pi} s^2
\\ && - \frac{20\alpha_s^2 \langle g_s^3 G^3 \rangle}{27} s \, ,
\label{eq:ope2pm}
\\
\nonumber \rho_{2^{--}}(s) &=& \frac{\alpha_s^3}{81 \pi} s^4 - \frac{5\alpha_s^2 \langle g_s^2 GG \rangle}{27} s^2 + \frac{15\alpha_s^3 \langle g_s^2 GG \rangle}{32\pi} s^2
\\ && - \frac{10\alpha_s^2 \langle g_s^3 G^3 \rangle}{27} s \, ,
\label{eq:ope2mm}
\\
\nonumber \rho_{3^{+-}}(s) &=& \frac{5\alpha_s^3}{2016 \pi} s^4 + \frac{\alpha_s^2 \langle g_s^2 GG \rangle}{16} s^2 - \frac{59\alpha_s^3 \langle g_s^2 GG \rangle}{512\pi} s^2
\\&& + \frac{13\alpha_s^2 \langle g_s^3 G^3 \rangle}{384} s \, ,
\\
\label{eq:ope3pm}
\nonumber \rho_{3^{--}}(s) &=& \frac{5\alpha_s^3}{2016 \pi} s^4 - \frac{\alpha_s^2 \langle g_s^2 GG \rangle}{16} s^2 - \frac{49\alpha_s^3 \langle g_s^2 GG \rangle}{1536\pi} s^2
\\ && + \frac{157\alpha_s^2 \langle g_s^3 G^3 \rangle}{3456} s \, .
\label{eq:ope3mm}
\end{eqnarray}
Especially, we find all the $D=8$ terms proportional to $\langle g_s^2 GG \rangle^2$ vanish, so the convergence of the above OPE series are quite good.

We need the strong coupling constant to perform numerical analyses~\cite{pdg}:
%
%%%%%%%%%%%%%%%%%%%%%%%%%%%%%%%%%%%%%%%%%%%%%%%%%%%%%%%%%%%%%%%%%%%%%%%%%%%%%%
\begin{equation}
\alpha_s(Q^2) = {4\pi \over 11 \ln(Q^2/\Lambda_{\rm QCD}^2)} \, ,
\end{equation}
%%%%%%%%%%%%%%%%%%%%%%%%%%%%%%%%%%%%%%%%%%%%%%%%%%%%%%%%%%%%%%%%%%%%%%%%%%%%%%
%
with the QCD scale at $\Lambda_{\rm QCD} = 300$~MeV.

We shall see that the odderon mass $M_X$ depends significantly on the gluon condensate $\langle g_s^2GG\rangle$. However, this parameter is still not well known, so we use altogether two sets of parameters:
\begin{itemize}

\item Parameter Set--I~\cite{Ioffe:2005ym}:
\begin{eqnarray}
\nonumber \langle \alpha_s GG\rangle &=& (0.005 \pm 0.004) \times \pi \mbox{ GeV}^4 \, ,
\\ \langle g_s^3G^3\rangle &=& \langle \alpha_s GG\rangle \times (8.2 \pm 1.0) \mbox{ GeV}^2 \, .
\label{eq:condensate1}
\end{eqnarray}

\item Parameter Set--II~\cite{Narison:2011xe,Narison:2018dcr}:
\begin{eqnarray}
\nonumber \langle \alpha_s GG\rangle &=& (6.35 \pm 0.35) \times 10^{-2} \mbox{ GeV}^4 \, ,
\\ \langle g_s^3G^3\rangle &=& \langle \alpha_s GG\rangle \times (8.2 \pm 1.0) \mbox{ GeV}^2 \, .
\label{eq:condensate2}
\end{eqnarray}

\end{itemize}

As shown in Eq.~(\ref{eq:LSR}), the mass $M_X$ depends on two free parameters, the threshold value $s_0$ and the Borel mass $M_B$. We use two criteria to determine the Borel window. Firstly, we investigate the convergence of OPE, which is the cornerstone of a reliable sum rule analysis. Because the $D=8$ term proportional to $\langle g_s^2 GG \rangle^2$ vanishes, this convergence is already quite good, while we further require the $\alpha_s^{n>3}$ terms $\alpha_s^3 \langle g_s^2 GG \rangle$ and $\alpha_s^2 \langle g_s^3 G^3 \rangle$ to be less than 5\%:
\begin{equation}
\mbox{CVG} \equiv \left|\frac{ \Pi^{\alpha_s^{n>3}}(s_0, M_B^2) }{ \Pi(s_0, M_B^2) }\right| \leq 5\% \, .
\label{eq:convergence}
\end{equation}
Secondly, we investigate the one-pole-dominance assumption, and require the pole contribution (PC) to be larger than 40\%:
\begin{equation}
\mbox{PC} \equiv \left|\frac{ \Pi(s_0, M_B^2) }{ \Pi(\infty, M_B^2) }\right| \geq 40\% \, .
\label{eq:pole}
\end{equation}
Altogether we can determine the Borel window for a fixed $s_0$. Then we change $s_0$ and redo the same procedures, so that we can find the lower bound of $s_0$.

Take the current $J_1^{\alpha\beta}$ as an example. When using the Parameter Set--I, there exist non-vanishing Borel windows as long as $s_0 > 14.1$~GeV$^2$, and the Borel window is determined to be $3.67$~GeV$^2 < M_B^2 < 4.13$~GeV$^2$ for $s_0 = 16$~GeV$^2$. Accordingly, we choose the working regions to be $14.0$~GeV$^2< s_0 < 18.0$~GeV$^2$ and $3.67$~GeV$^2 < M_B^2 < 4.13$~GeV$^2$, and calculate the mass of the $1^{+-}$ odderon to be
\begin{equation}
M_{X_{1^{+-}}} = 2.87^{+0.17}_{-0.20}{\rm~GeV} \, .
\end{equation}
Its uncertainty is quite large, mainly coming from the uncertainty of the gluon condensate $\langle g_s^2GG\rangle$ given in Eqs.~(\ref{eq:condensate1}).

Similarly, we use the currents $J_{1/2/3}^{\alpha_1\cdots\alpha_J,\beta_1\cdots\beta_J}$ and $\tilde J_{1/2/3}^{\alpha_1\cdots\alpha_J,\beta_1\cdots\beta_J}$ to perform numerical analyses, and calculate masses of the $J^{PC} = 1/2/3^{\pm-}$ odderons. The obtained results are summarized in Table~\ref{tab:mass}. Note that the OPE convergence is sometimes so good that the lower bound of $M_B$ can not be well determined, for which cases we need to properly choose $s_0$ according to their partner states, {\it e.g.}, see the result of $X_{2^{+-}}$ using the Parameter Set--II. We clearly see from Table~\ref{tab:mass} that our QCD sum rule results depend significantly on the gluon condensate $\langle g_s^2GG\rangle$, which is currently not well known and still waiting to be clarified.

\begin{table*}[hptb]
\begin{center}
\renewcommand{\arraystretch}{1.45}
\caption{Masses of the $J^{PC} = 1/2/3^{\pm-}$ odderons, extracted from the currents $J_{1/2/3}^{\alpha_1\cdots\alpha_J,\beta_1\cdots\beta_J}$ and $\tilde J_{1/2/3}^{\alpha_1\cdots\alpha_J,\beta_1\cdots\beta_J}$. In the Parameter Set--I we choose the gluon condensate to be $\langle \alpha_s GG\rangle = (0.005 \pm 0.004) \times \pi$~GeV$^4$~\cite{Ioffe:2005ym}, and in the Parameter Set--II we choose it to be $\langle \alpha_s GG\rangle = (6.35 \pm 0.35) \times 10^{-2}$~GeV$^4$~\cite{Narison:2018dcr}.}
\begin{tabular}{c | c | c | c | c | c | c | c}
\hline\hline
~~~~~ & \multirow{2}{*}{~~Odderon~~} & \multirow{2}{*}{~~~~~~Current~~~~~~} & \multirow{2}{*}{~$s_0^{min}~[{\rm GeV}^2]$~} & \multicolumn{2}{c|}{Working Regions} & \multirow{2}{*}{~~Pole~[\%]~~} & \multirow{2}{*}{~~Mass~[GeV]~~}
\\ \cline{5-6} & & & & ~~$s_0~[{\rm GeV}^2]$~~ & ~~$M_B^2~[{\rm GeV}^2]$~~ &
\\ \hline \hline
\multirow{6}{*}{\rotatebox{90}{Parameter Set--I}}
& $X_{1^{+-}}$   &   $J_1^{\alpha\beta}$                                            & 14.1 & $16.0\pm2.0$ & $3.67$--$4.13$ & $40$--$50$ & $2.87^{+0.17}_{-0.20}$
\\
& $X_{2^{+-}}$   &   $J_2^{\alpha_1\alpha_2,\beta_1\beta_2}$                        & 10.6 & $16.0\pm2.0$ & $2.76$--$4.07$ & $40$--$73$ & $2.85^{+0.16}_{-0.20}$
\\
& $X_{3^{+-}}$   &   $J_3^{\alpha_1\alpha_2\alpha_3,\beta_1\beta_2\beta_3}$         & 8.9  & $16.0\pm2.0$ & $2.60$--$4.23$ & $40$--$81$ & $2.78^{+0.18}_{-0.23}$
\\
& $X_{1^{--}}$   &   $\tilde J_1^{\alpha\beta}$                                     & 15.1 & $17.0\pm2.0$ & $2.93$--$3.52$ & $40$--$54$ & $3.29^{+1.49}_{-0.32}$
\\
& $X_{2^{--}}$   &   $\tilde J_2^{\alpha_1\alpha_2,\beta_1\beta_2}$                 & 15.3 & $17.0\pm2.0$ & $3.29$--$3.74$ & $40$--$50$ & $3.16^{+0.33}_{-0.23}$
\\
& $X_{3^{--}}$   &   $\tilde J_3^{\alpha_1\alpha_2\alpha_3,\beta_1\beta_2\beta_3}$  & 15.0 & $17.0\pm2.0$ & $2.55$--$3.36$ & $40$--$58$ & $3.47^{+~\,?}_{-0.50}$
\\ \hline \hline
\multirow{6}{*}{\rotatebox{90}{Parameter Set--II}}
& $X_{1^{+-}}$   &   $J_1^{\alpha\beta}$                                            & 19.0 & $21.0\pm2.0$ & $5.38$--$5.88$ & $40$--$47$ & $3.27^{+0.15}_{-0.17}$
\\
& $X_{2^{+-}}$   &   $J_2^{\alpha_1\alpha_2,\beta_1\beta_2}$                        & 12.6 & $21.5\pm2.0$ & $3.64$--$5.84$ & $40$--$79$ & $3.28^{+0.14}_{-0.19}$
\\
& $X_{3^{+-}}$   &   $J_3^{\alpha_1\alpha_2\alpha_3,\beta_1\beta_2\beta_3}$         & 19.8 & $22.0\pm2.0$ & $5.77$--$6.27$ & $40$--$47$ & $3.30^{+0.15}_{-0.17}$
\\
& $X_{1^{--}}$   &   $\tilde J_1^{\alpha\beta}$                                     & 32.6 & $35.0\pm3.0$ & $6.05$--$6.93$ & $40$--$49$ & $5.05^{+0.17}_{-0.14}$
\\
& $X_{2^{--}}$   &   $\tilde J_2^{\alpha_1\alpha_2,\beta_1\beta_2}$                 & 29.1 & $36.0\pm3.0$ & $5.96$--$7.93$ & $40$--$63$ & $4.72^{+0.15}_{-0.17}$
\\
& $X_{3^{--}}$   &   $\tilde J_3^{\alpha_1\alpha_2\alpha_3,\beta_1\beta_2\beta_3}$  & 34.5 & $37.0\pm3.0$ & $5.90$--$7.07$ & $40$--$50$ & $5.45^{+0.32}_{-0.21}$
\\ \hline\hline
\end{tabular}
\label{tab:mass}
\end{center}
\end{table*}

$\\$
{\it Summary and discussions.}-----
In this letter we apply the method of QCD sum rules to study the odderon as a three-gluon bound state. There may exist six lowest-lying odderons with the quantum numbers $J^{PC} = 1/2/3^{\pm-}$. We systematically construct their interpolating currents using the gluon field strength tensors $G^a_{\mu\nu}$ and $\tilde G^a_{\mu\nu}$. All these currents have $N=2J$ Lorentz indices with certain symmetries, so that they couple to both the positive- and negative-parity odderons, which need to be further separated at the hadron level. The construction of such currents is quite general and may be applied in fields other than hadron physics.

We construct altogether six relativistic low-dimension odderon currents with the quantum numbers $J^{PC} = 1/2/3^{\pm-}$. We use them to perform QCD sum rule analyses, and calculate masses of the $J^{PC} = 1/2/3^{\pm-}$ odderons. The results are summarized in Table~\ref{tab:mass}, sometimes with quite large uncertainties coming from the gluon condensates $\langle g_s^2GG\rangle$ and $\langle g_s^3G^3\rangle$. It is interesting to compare our results with the Lattice QCD results~\cite{Chen:2005mg,Mathieu:2008me,Meyer:2004gx,Gregory:2012hu} obtained using non-relativistic odderon operators, as given in Table~\ref{tab:comparison}.

\begin{table*}[hptb]
\begin{center}
\renewcommand{\arraystretch}{1.45}
\caption{Masses of the $J^{PC} = 1/2/3^{\pm-}$ odderons, in units of GeV. Our QCD sum rule results are obtained using the Parameter Set--I and Set--II, and the Lattice QCD results are taken from Refs.~\cite{Chen:2005mg,Mathieu:2008me,Meyer:2004gx} (quenched) and Ref.~\cite{Gregory:2012hu} (unquenched).}
\begin{tabular}{c | c | c | c | c | c | c}
\hline\hline
~~Odderon~~  & ~~~~Set--I~~~~         & ~~~~Set--II~~~~        & ~~~~~~~~Ref.~\cite{Chen:2005mg}~~~~~~~~ & ~~~~~~~~Ref.~\cite{Mathieu:2008me}~~~~~~~~ & ~~~~~~~~Ref.~\cite{Meyer:2004gx}~~~~~~~~  & ~~~Ref.~\cite{Gregory:2012hu}~~~
\\ \hline
$X_{1^{+-}}$ & $2.87^{+0.17}_{-0.20}$ & $3.27^{+0.15}_{-0.17}$ & $2.98 \pm 0.03 \pm 0.14$                & $2.94 \pm 0.03 \pm 0.14$                   & $2.67 \pm 0.07 \pm 0.12$                  & $3.27 \pm 0.34$
\\
$X_{2^{+-}}$ & $2.85^{+0.16}_{-0.20}$ & $3.28^{+0.14}_{-0.19}$ & $4.23 \pm 0.05 \pm 0.20$                & $4.14 \pm 0.05 \pm 0.20$                   & --                                        & --
\\
$X_{3^{+-}}$ & $2.78^{+0.18}_{-0.23}$ & $3.30^{+0.15}_{-0.17}$ & $3.60 \pm 0.04 \pm 0.17$                & $3.55 \pm 0.04 \pm 0.17$                   & $3.27 \pm 0.09 \pm 0.15$                  & $3.85 \pm 0.35$
\\
$X_{1^{--}}$ & $3.29^{+1.49}_{-0.32}$ & $5.05^{+0.17}_{-0.14}$ & $3.83 \pm 0.04 \pm 0.19$                & $3.85 \pm 0.05 \pm 0.19$                   & $3.24 \pm 0.33 \pm 0.15$                  & --
\\
$X_{2^{--}}$ & $3.16^{+0.33}_{-0.23}$ & $4.72^{+0.15}_{-0.17}$ & $4.01 \pm 0.05 \pm 0.20$                & $3.93 \pm 0.04 \pm 0.19$                   & $3.66 \pm 0.13 \pm 0.17$                  & $4.59 \pm 0.74$
\\
$X_{3^{--}}$ & $3.47^{+~\,?}_{-0.50}$ & $5.45^{+0.32}_{-0.21}$ & $4.20 \pm 0.05 \pm 0.20$                & $4.13 \pm 0.09 \pm 0.20$                   & $4.33 \pm 0.26 \pm 0.20$                  & --
\\ \hline \hline
\end{tabular}
\label{tab:comparison}
\end{center}
\end{table*}

From the above comparison, we can see how we poorly understand the odderon. Recall that there is currently no definite experimental evidence for the existence of any glueball yet, we quickly realize how important is the evidence of the odderon exchange recently observed by D0 and TOTEM~\cite{Abazov:2020rus}. Since this is still an indirect evidence, we propose to directly search for the odderon at LHC.

From the viewpoint of quark model, the odderon can decay after exciting three quark-antiquark pairs, and recombine into three mesons. Generally speaking, its width can be quite large, preventing it to be easily observed. We use $P$ and $V$ to denote the light vector and pseudoscalar mesons respectively, and its possible decay patterns are:
\begin{eqnarray}
\nonumber 1^{--} &\to& ~~~~\,VPP,VVP,VVV~~~~~(S\mbox{-wave}) \, ,
\\
\nonumber 1^{+-} &\to& PPP,VPP,VVP,VVV~(P\mbox{-wave}) \, ,
\\
\nonumber 2^{--} &\to& ~~~~~~~~\,VVP,VVV~~~~~~~~~(S\mbox{-wave}) \, ,
\\
\nonumber 2^{+-} &\to& ~~~~VPP,VVP,VVV~~~~~(P\mbox{-wave}) \, ,
\\
\nonumber 3^{--} &\to& ~~~~~~~~~~~~\,VVV~~~~~~~~~~~~~(S\mbox{-wave}) \, ,
\\
\nonumber 3^{+-} &\to& ~~~~~~~~VVP,VVV~~~~~~~~~(P\mbox{-wave}) \, .
\end{eqnarray}
Due to their limited decay patterns, the spin-3 odderons have relatively smaller widths probably, and so we propose to search for them in their $VVV$ and $VVP$ decay channels directly at LHC.

%
%=====================================================================================
%=====================================================================================
%=====================================================================================
\section*{Acknowledgments}
%=====================================================================================
%=====================================================================================
%=====================================================================================
%

SLZ is grateful to Shou-Hua Zhu for helpful discussions.
This project is supported by
the National Natural Science Foundation of China under Grant No. 11722540, No.~11975033, No. 12075019, and No.~12070131001,
the National Key Research and Development Program of China (2020YFA0406400),
and
the Fundamental Research Funds for the Central Universities.

\end{document}